# A SECURE SERVICE PROVISIONING FRAMEWORK FOR CYBER PHYSICAL CLOUD COMPUTING SYSTEMS

Anees Ara, Mznah Al-Rodhaan, Yuan Tian, Abdullah Al-Dhelaan

Department of Computer Science, College of Computers and Information Sciences, King Saud University, Riyadh, Saudi Arabia

## ABSTRACT

*Cyber physical systems (CPS) are mission critical systems engineered by combination of cyber and physical systems respectively. These systems are tightly coupled, resource constrained systems and have dynamic real time applications. Due to the limitation of resources, and in order to improve the efficiency of the CPS systems, they are combined with cloud computing architecture, and are called as Cyber Physical Cloud Computing Systems (CPCCS). These CPCCS have critical care applications where security of the systems is a major concern. Therefore, we propose a Secure Service provisioning architecture for Cyber Physical Cloud Computing Systems (CPCCS), which includes the combination of technologies such as CPS, Cloud Computing and Wireless Sensor Networks. In addition to this, we also highlight various threats/attacks; security requirements and mechanisms that are applicable to CPCCS at different layers and propose two security models that can be adapted in a layered architectural format.*

## 1. INTRODUCTION

Cyber Physical Systems are large scaled, closely integrated and resource constrained, collection of distributed cyber and physical systems respectively [2]. In CPS, the physical systems and its processes are monitored, coordinated and controlled by the computation and communication cores. All these physical devices have limited resources to satisfy complex communication & computational tasks needed by critical applications. And hence these limitations were overcome by the introduction of the concept of Service oriented architecture (SOA), of cloud computing systems [22], being applied to CPS. Consequently, this improves the interoperability, flexibility and efficiency of the CPS [7, 14].

In 2013, The National Institute of standards and technology (NIST) have defined Cyber Physical Cloud Computing (CPCC) architectural framework as "a system environment that can rapidly build, modify and provision auto-scale cyber-physical systems, composed of a set of cloud computing based sensor, processing, control, and data services" [6]. These systems provide mission critical services such as 1) Smart Health: medical devices & systems, automated pervasive health; 2) Smart Transportation: Air CPS, Intelligent transport systems ITS, Unmanned Aerial Vehicles UAV; 3) Smart Power grid: electrical power grid; Navigation & rescue applications; 4) Social networking and gaming [6, 11, 13, 29]. Since all these are mission critical applications, and hence system security & privacy becomes one of the major requirement and research challenge needed by CPCCS [6, 13].

In literature, there are various approaches published in the direction of combining service oriented architecture (cloud computing) and CPS [7, 8, 12, 15], in which the CPS were considered as service models; platform as service (PaaS) and Infrastructure as service (IaaS). In addition to these there are application specific approaches, where more prominence was given to the amalgamation of technologies such as Cyber Physical System, Cloud Computing and Wireless sensor networks [9, 10, 11, 18], but none of them focussed on security and privacy related issues. Hence by taking the motivation, from the above approaches, combining these technologies (CPS, Cloud computing and WSN) and also by considering various security issues related to them, we propose a secure service provisioning framework for CPCCS.

Our contributions in this paper include the following:

1) We proposed A Secure Service Provisioning architecture for Cyber Physical Cloud Computing Systems, through which, the immense data from the connected physical sensing devices (such as WSN) can be securely provided as service provisioning sub systems (such as XaaS) of cloud computing architecture [7, 8, 9, 10].
2) We surveyed various threats/attacks, security requirements and mechanisms applicable to CPCCS at different layers.
3) We propose two security models namely horizontal security model and vertical security model that can be adapted in a layered architectural format in CPCCS.
The rest of the paper is organized as follows: Section 2 discusses the related works in securing CPS. Section 3 proposes Secure Service Provisioning architecture for CPCCS, which includes the classification of sub systems been provided as services in a secure manner. Section 4 summarizes the various security challenges and requirements related to CPCCS and propose two security models related to it. Section 5 concludes the whole paper.

## 2. RELATED WORKS

The CPS, Internet of things (IoT) and Sensor Cloud are built on the same realms of cyber-physical systems plethora supported by the underlying M2M communication paradigm. Security in these areas has been a prominent area of research in recent years.

Some of the proposed security architectures were application specific in CPS. Ahmed Lounis et al [19] proposed a secure cloud based architecture for e-Health wireless sensor networks. Through the combination of attribute-based encryption and symmetric cryptography they try to eliminate the threats & risk of outsourcing the medical data to cloud computing systems. Yin-Jun Chen et al [17] has developed the CPS integrated framework for monitoring cultural heritage. They used RSA algorithm and fuzzy theory in detecting abnormalities or intrusions in the communication network. In addition to these HSCloud [16], a homeland security architecture was developed to monitor the transportation of dangerous goods (TGS) by using Cloud enabled virtual environment (CLEVER) and a Virtual pervasive element (VPE).

Some other works include the solutions applied to a particular security issue such as the CRBAC (Context-aware Role Based Access Control), a framework for Cyber physical cloud, proposed by Hiray et al [5], which can be used for controlling, tracing and providing an authorized access to system resources. Also finite state model and hidden Markov chains were adopted to identify the multi-stage attacks and to prevent real damage in CPS[4].Besides, the adoption of FPGAs in protecting Cyber Physical Infrastructures was suggested to improve security and robustness of heterogeneous networks [3]. Chonka et al developed a defence system called Pre-Decision, Advance Decision, Learning System (ENDER) for detecting and mitigating HX(HTTP and XML)-DoS attacks against cloud web services in Cyber physical cloud systems[20, 30].

In addition, Chris et al [21] presented game theoretic formulations of attack and defence aspects of CPS under different cost - benefit functions and different budgets of the attacker and defender. Quanyan Zhu [25] developed a game-theoretic framework for addressing security and resilience problems at multiple layers of the cyber-physical systems which includes robust and resilient control, secure network routing and management of information security and smart grid energy systems.

## 3. THE PROPOSED SECURE SERVICE PROVISIONING FRAMEWORK FOR CPCCS

CPCCS are basically a CPS that combines sensor networks with cloud computing systems. The proposed framework is one such combination of sensing devices and cloud computing systems being provided in a secure way. In the proposed framework the sensor networks are being provided as service provisioning sub systems (XaaS) of cloud computing paradigm.

The proposed framework is a combination of resource oriented architecture [28] and service oriented architectures [27], in which the utilities of the multiple physical sensing networks are provisioned as services through cloud computing systems in a secure way. That is the immense critical data which is continuously sensed by various wireless sensor networks are combined and securely provisioned as service modules through commercial application systems in the cloud environment.

### 3.1. Basic Components

- *Wireless Sensor Networks (WSN):* It is a network that comprises of the low power sensing devices called as nodes and a base station/sink which is relatively stronger in power and storage. These nodes with base station are deployed in a physical region to sense a particular entity of environment such as temperature, pressure, humidity etc.
- *Gateway (proxy):* It is a proxy server owned by the WSN owner who is responsible for data aggregation and dissemination to different applications.
- *Virtual Sensor Networks (VSN):* It is a seamless grouping of WSN's in a virtual environment like an overlay.
- *Application Management Unit (AMU):* It comprises of Interface agent, service API's Service list. The general functionality of AMU is to manage the service request which is posed by the client/user through an interface agent. The interface agent looks for the request in the service list and responses back to the client about its availability and if in case of its absence it integrates or develops new Service API's according the user requirement specification.
- *Management/Monitoring Unit (MU):* The responsibilities of MU are to manage and monitor the different virtual components of the system through resource management, load balancing, network resource management, real time scheduling and end – end performance evaluation metrics.
- *Supervisory Agent:* The main responsibility of this agent is to supervise the pub/sub unit, where the user requests are sent through interface unit for the subscription of a service. The supervisory agent by using event matching & processing unit tries to find if the request can be matched by searching for the service in service registry. If the match is found the data in response to the request is send back.
- *Security Unit (SU):* The basic operations of the SU are to provide security at various levels from the time a user puts in a request till the response is fed back. That is an end to end security which is provided from physical layer to application layer. It includes the following components as:

- *Identity Management Unit (IMU)* who is responsible for authentication and authorization of the various assets in the system.
- *Access Control Unit (ACU)* with the help of Access Control List manages/grants a user/client, access permission/rights on a particular object (here known as service) and the operations performed by it.
- *Storage security* is responsible for applying the security mechanisms on the virtual database storage servers.
- *Service level Agreements (SLA):* These are the agreements made between the cloud service provider and the user/client in this system, which includes agreements made by both the parties about the list of services, metrics, audit mechanisms, security, encryption, privacy, data retention, regulatory compliance and transparencies to be provided by Cloud service providers to the users.
- *Policy Management*: A proper policy management has to be done between multiple WSN owners by the CPS cloud service provider for the information been provided by them.
- *Security audit:* Since the system comprises of cyber and physical components, the security audit should be processed at various levels i.e., information security audit, network security audit & physical security audit.

### 3.2. Types of service provisioning sub-systems

In this proposed architecture (Figure.1) we classify the type of sub systems that can be provisioned as (XaaS) services in the CPCCS with the help of following three cases:

**Case 1.** Publish/Subscribe systems as SaaS

SaaS is the software that is deployed over the internet. When a client/user request for data, the cloud service provider (CSP) using SaaS applications subscribes to pub-sub system and the service for which user requested is searched in the service registry, if the match is found the data as a service is published to the user.

In our proposed framework, the users are at the application layer of the architecture where they interact with the system using the interface agent, and place a subscription request to the system. The supervisory agent then looks for the match in the service registry using the event matching /processing unit. If the match is found the data is published. Depending on the type of request that is, If the user/subscriber request for raw data then according to the availability of data at DB storage server or from the WSN owner, the XML data is directly provided in a secure encrypted form through service API's.

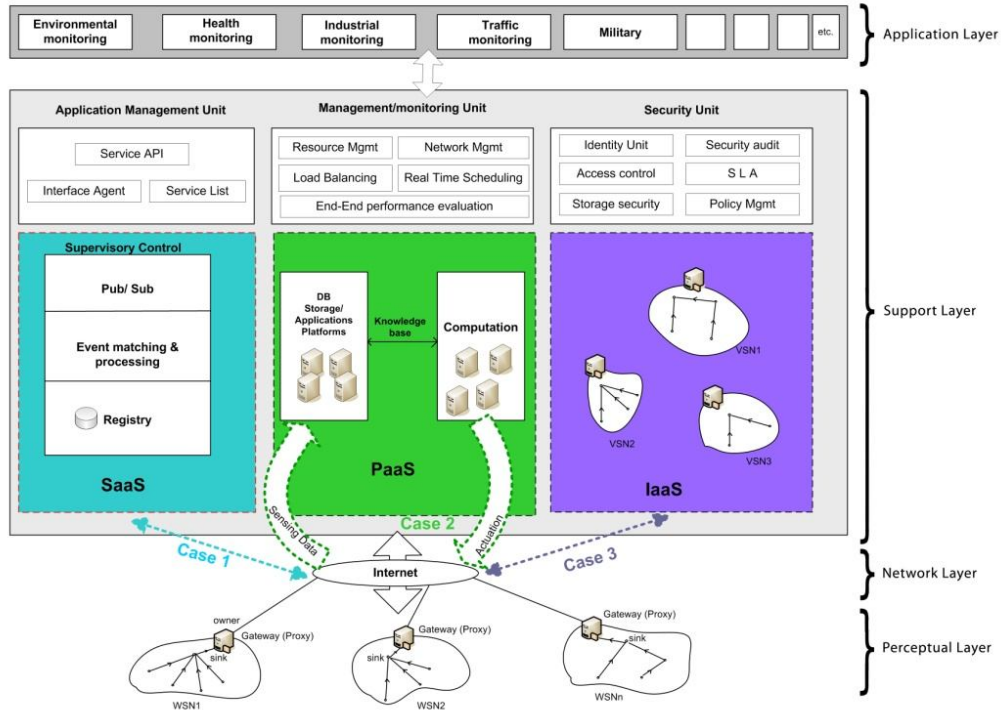

Figure 1. Proposed Secure Service Provisioning Framework for CPCCS: Case (1). Pub/Sub systems as SaaS, Case (2). Sensing and Actuation Systems as PaaS and Case (3). Virtual Sensor Networks as IaaS.

**Case 2.** Sensing and Actuation Systems as PaaS

These are the cyber physical systems where the sensor nodes form the physical sensing systems, the cloud storage and computation platform forms the cyber back bone.

These systems are used for controlling and monitoring the critical environments like industrial monitoring systems, intelligent transport systems, health monitoring systems, SCADA etc. The case (2) in the architecture discusses about these type of the systems where physical entities in a particular area can be sensed using the resource constraint sensor nodes and the data is periodically send to the cloud storage using the secure communication protocols. The computation on the sensed data is performed on the cloud servers. Based on the knowledge driven from the computation; the results are sent back so that the actuation can take place.

**Case 3.** Virtual Sensor networks as IaaS

The virtual sensor networks (VSN) are the overlays on the wireless sensor networks. Each VSN is actually a mapping of sensor nodes to the virtual machines or the virtual instances. According to the application requirement posed by the client/users, the corresponding data can be sent from multiple sensors which are mapped onto the virtual machine or virtual instance. The Case (3) in our architecture discusses about this kind of system. As we know that the deployment of each WSN for only one sensing application is very cost effective. It is also found that there are overlapping sensor nodes from two different sensor networks deployed in common region. Therefore, through the virtual sensor network environment, a seamless grouping of multiple sensor networks can be done on the virtual instance and correlations drawn data from them can be of great help to various applications.

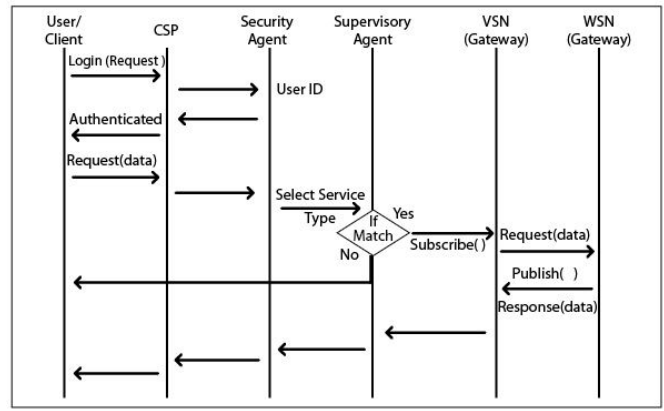

Figure 2. Workflow diagram for the proposed secure service provisioning framework of CPCCS.

The general workflow scenario for CPCCS from the application layer to the physical layer is described in Figure 2. It begins when a user/client wants to access the system, he first puts a login request to the cloud service provider (CSP) and the request is forwarded to the security agent. The Security Agent generates a UserID and the user is authenticated. An authenticate user makes request for the required data, which gets forwarded from CSP to the supervisory agent. The supervisory agent by using event matching & processing unit tries to find if the request can be matched and found in the service registry. If the match is found, then according to the subscribers request the data will be published as a response from the underlying physical network (WSN).

## 4. SECURING CYBER PHYSICAL CLOUD COMPUTING SYSTEMS

The integration of cyber (cloud computing and other computing devices) and physical (physical sensing devices) components with the heterogeneous networks in a cyber-physical cloud computing system introduces different types of vulnerabilities that include attacks on physical Space, attacks on cyber space and interception, replacement or removal of information from the communication channels[26]. The major challenges for securing Cyber Physical systems constitute in understanding the threats and attacks, identifying the features of CPS and their differences from traditional IT security and applicability of available security mechanisms in CPS [1].

In Table 1, the threats and attacks, security requirements and security mechanisms applicable according to the components at each layer in Cyber Physical Cloud Computing Systems are summarized based on [23, 24].

### 4.1. Proposed Security Models

Using layered architecture format of the security framework discussed in section 3, we propose two types of security models as follows:

#### 4.1.1. Horizontal Security Model

The horizontal security model consists of horizontal decomposition of components in CPCCS. It includes Cloud Computing infrastructure, Communication links and Wireless Sensor Network in a top-down hierarchy as shown in Figure 3. The horizontal security model decomposes CPCCS system in a horizontal manner and it involves securing each level as follows:

- Secure in-network data aggregation at WSN

Table 1. Security Attacks, Requirements and Mechanisms applicable to CPCCS

| Layers | Components | Threats/Attacks | Security requirements | Security management mechanisms |
|---|---|---|---|---|
| Application layer | Personalized information service, intelligent transportation, environmental monitoring | DoS, DDoS, Cross site scripting (XSS), Injection attacks, buffer overflow, Session hijacking, Security misconfiguration, Failure to restrict URL access. | Authentication and key agreement, privacy protection, security education and management | Access control management, security management, privacy protection management |
| Support layer | Cloud computing, intelligent computing | Data Breaches, Data Loss, Account Hijacking, Insecure APIs, Denial of Service (DoS), Malicious Insiders, Abuse of Cloud Services, Insufficient Due Diligence and Shared Technology Issues | Secure Multiparty computation, secure cloud computing, Anti-Virus | Behavior entities certification, Data metric, Key generation and distribution, security & computation |
| Network layer | Internet mobile communication network, Satellite nets, network infrastructure and communication protocols | Eavesdropping, Data Modification, Identity Spoofing (IP Address Spoofing), Password-Based Attacks, Denial-of-Service Attack, Man-in-the-Middle Attack, Compromised-Key Attack and Sniffer Attack | Identity Authentication, Anti DDOS, Encryption mechanism, Communication security | User privacy, Data encryption, Data integrity, Multicast Security, Entity Authentication, Access security |
| Perceptual layer | RFID reader, Sensor, GPS | DDoS due to jamming, Physical devices destruction, obstruction, manipulation, or malfunction of physical assets | Light weight encryption technology, protecting sensor data, key agreement | Intrusion detection, wireless encryption, key management, reputation evaluation, secure routing, Distributed authentication |

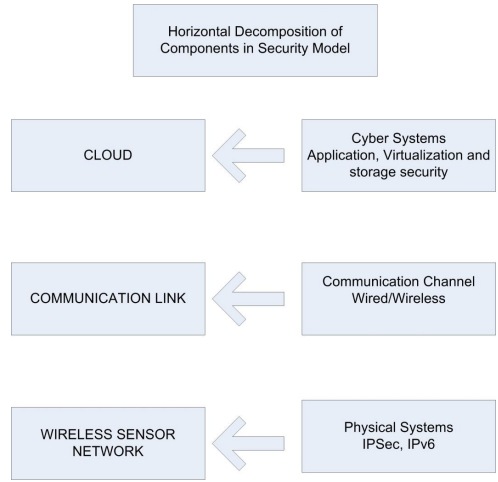

Figure 3. Proposed Horizontal security model for CPCCS

- Secure data aggregation and dissemination from multiple WSN's, through secure communication channel (internet)
- Ensuring an overall security at cloud as storage security, computational security, virtual machines security and application management security.

### 4.1.2. Vertical Security Model

The vertical security model consists of vertical decomposition of layers in CPCCS. It includes Application Layer, XaaS (SaaS/PaaS/IaaS), Communication links and Wireless Sensor Networks in a top-down hierarchy as explained in Figure 4. The vertical decomposition ensures that the data is collected in a vertical pattern as follows:

- Secure in-network data aggregation at WSN
- Secure data aggregation and dissemination from multiple WSN's, through secure communication channel to the cloud.

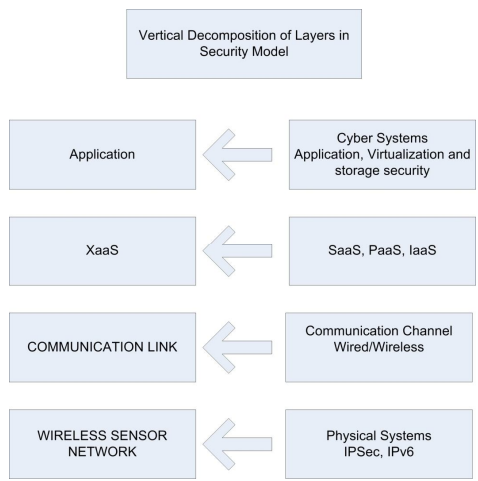

Figure 4. Proposed Vertical security model for CPCCS

- Through a secure service provisioning system in cloud a service model (SaaS/PaaS/IaaS) been to be selected according to the application/client requirement.
- Using the best web security mechanisms and following the application security protocols, the data is send to the Client as requested.

## 5. CONCLUSION

In this paper we have proposed a novel security architectural framework for CPCCS with different service provisioning sub-systems, which includes three cases, namely 1) Publish/Subscribe systems as Software as service (SaaS), 2) Sensing and Actuation Systems as Platform as service (PaaS) and 3) Virtual Sensor networks as Infrastructure as service (IaaS). We have also surveyed various components, threats and attacks, security requirements and security mechanisms applicable at different layers of CPCCS. We also proposed two security models namely Horizontal security model and Vertical security model, based on the integration criteria of CPCCS. As for the future work we focus on studying various cryptographic techniques that can be applied to secure real time CPCCS. We intend to secure CPCCS using a bottom up approach, by developing algorithms that can be implemented to perform secure data aggregation in physical sensing networks in CPCCS.

## ACKNOWLEDGMENT

This project was supported by NSTIP strategic technologies program number 13-INF2080-02 in the Kingdom of Saudi Arabia

International Journal of Distributed and Parallel Systems (IJDPS) Vol.6, No.1, January 2015

## Authors Biography

**Anees Ara** received her BSc in Computer Science and MSc in Mathematics with Computer Science from Osmania University, Hyderabad, India in 2005 and 2007 respectively. Currently, she is a Ph.D. candidate in the Department of Computer Science at College of Computers and Information Science, King Saud University, Riyadh. Her research interest include IoT, sensor networks, cloud computing and security in pervasive and ubiquitous computing.

**Mznah Al-Rodhaan** has received her BS in Computer Applications (Hon) and MS in Computer Science both from King Saud University on 1999 and 2003 respectively. In 2009, she received her Ph.D. in Computer Science from University of Glasgow in Scotland, UK. She is currently working as the Vice Chair of the Computer Science Department in College of Computer & Information Sciences, King Saud University, Riyadh, Saudi Arabia. Moreover, she has served in the editorial boards for some journals such as the Ad Hoc journal (Elsevier) and has participated in several international conferences. Her current research interest includes: Mobile Ad Hoc Networks, Wireless Sensor Networks, Multimedia Sensor networks, Cognitive Networks, and Network Security.

**Yuan Tian** has received her master and PhD degree from Kyung Hee University and she is currently working as Assistant Professor at College of Computer and Information Sciences, King Saud University, Kingdom of Saudi Arabia. She is member of technical committees of several international conferences. In addition, she is an active reviewer of many international journals. Her research interests are broadly divided into privacy and security, which are related to cloud computing, bioinformatics, multimedia, cryptograph, smart environment, and big data.

**Abdullah Al-Dhelaan**, has received BS in Statistics (Hon) from King Saud University, on 1982, and the MS and Ph.D. in Computer Science from Oregon State University on 1986 and 1989 respectively. He is currently the Vice Dean for Academic Affairs, Deanship of Graduate Studies and a Professor of Computer Science, King Saud University, Riyadh, Saudi Arabia. He has guest edited several special issues for the Telecommunication Journal (Springer), and the International Journal for Computers and their applications (ISCA). Moreover, he is currently on the editorial boards of several 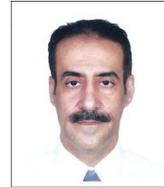 journals and the organizing committees for several reputable international conferences. His current research interest includes: Mobile Ad Hoc Networks, Sensor Networks, Cognitive Networks, Network Security, Image Processing, and High Performance Computing.